\documentclass[journal]{IEEEtran}
\usepackage{amsmath,amsfonts}
\usepackage{algorithm}
\usepackage{array}
\usepackage{caption}
\usepackage[font=footnotesize]{subcaption}
\usepackage{textcomp}
\usepackage{stfloats}
\usepackage{url}
\usepackage{verbatim}
\usepackage{graphicx}
\usepackage{tikz}
\usepackage{hyperref}
\usepackage{xurl}
\usepackage{booktabs}
\usepackage{cleveref}
\crefname{apdx}{appendix}{appendices}
\usepackage{microtype}
\usepackage{algpseudocode}
\usepackage{enumitem}

\hypersetup{
    colorlinks=true,
    linkcolor=blue,
    urlcolor=blue,
    citecolor=blue,
    pdftitle={Multi-objective Bayesian optimisation for design of Pareto-optimal current drive profiles in STEP},
    pdfpagemode=FullScreen,
}

\newcommand{\printbibliography}{\bibliographystyle{IEEEtran}\bibliography{main}}
\let\textcite\cite

\newtheorem{definition}{Definition}

\begin{document}
\title{Multi-objective Bayesian optimisation for design of Pareto-optimal current drive profiles in STEP}
\author{Theodore Brown, Stephen Marsden, Vignesh Gopakumar, Alexander Terenin, Hong Ge, and Francis Casson
\thanks{Submitted 4\textsuperscript{th} October 2023; revised version submitted 8\textsuperscript{th} February 2024.}%
\thanks{Theodore Brown was with the Department of Engineering, University of Cambridge, and is now with University College London and the United Kingdom Atomic Energy Authority (e-mail: \textsc{\href{mailto:theo.brown@ukaea.uk}{theo.brown@ukaea.uk}}). 
Stephen Marsden, Vignesh Gopakumar, and Francis Casson are with the United Kingdom Atomic Energy Authority.
Hong Ge is with the Department of Engineering, University of Cambridge. 
Alexander Terenin was with the Department of Engineering, University of Cambridge, and is now with the Center for Data Science for Enterprise and Society, Cornell University.}%
}

\markboth{Submitted to IEEE Transactions on Plasma Science, October 2023 [preprint]}{Brown \MakeLowercase{\emph{et al.}}: Multi-objective Bayesian optimisation for design of Pareto-optimal current drive}

\maketitle

\begin{abstract}
The safety factor profile is a key property in determining the stability of tokamak plasmas.
To design the safety factor profile in the United Kingdom's proposed Spherical Tokamak for Energy Production (STEP), we apply multi-objective Bayesian optimisation to design electron-cyclotron heating profiles.
\emph{Bayesian optimisation} is an iterative machine learning technique that uses an uncertainty-aware predictive model to choose the next designs to evaluate based on the data gathered during optimisation.
By taking a \emph{multi-objective} approach, the optimiser generates sets of solutions that represent optimal tradeoffs between objectives, enabling decision makers to understand the compromises made in each design.
The solutions from our method score higher than those generated in previous work by a genetic algorithm; however, the key result is that our method returns a \emph{purposefully diverse range} of optimal solutions, providing more information to tokamak designers without incurring additional computational cost.
\end{abstract}

\begin{IEEEkeywords}
Fusion reactor design, optimisation methods, Gaussian processes, machine learning
\end{IEEEkeywords}

\section{Introduction}
The starting point in the development of high-performance tokamak scenarios is the selection of the desired plasma properties during the main phase of operation, known as the flat-top operating point (FTOP).
The chosen flat-top scenario determines not only the energy-generating performance, but also the controllability and stability of the plasma \cite{walker2020}.

The design space for tokamak devices is very large, involving the interaction of many hundreds of parameters and many different physical regimes: consequently, it is very difficult to find designs that meet the criteria and constraints for a successful device. 
Exploring the space is a particular challenge for the UK's flagship Spherical Tokamak for Energy Production (STEP), as the novelty of spherical tokamak technology and power-generating plasmas introduces significant uncertainty into the design process.
In such a vast and changing space, it is important to be able to quantify the tradeoffs between competing objectives and quickly observe the effects of reformulating or adjusting the FTOP specification.

FTOP candidates for STEP are simulated using JETTO, a modelling code that iteratively solves coupled magnetic equilibrium and plasma transport equations \cite{cenacchi1988}.
On state-of-the-art high performance computing, JETTO takes around 3 hours to simulate a new FTOP\footnote{Benchmarks performed on a single 2.20GHz Intel Xeon Platinum 8276 CPU with 3380 MiB of RAM, hosted by the Cambridge Service for Data-driven Discovery. As the main cost is evolving the temporal rather than the spatial grid, this particular problem has minimal benefit from further parallelisation.}, where the plasma is simulated for several resistive timescales until it converges to a steady state.
As scenario optimisation typically requires hundreds of JETTO runs, the computational cost involved is significant.

The major contribution of this paper is the application of an improved method for resolving the tradeoffs between plasma design objectives.
We show that taking a direct multi-objective approach results in interpretable solutions to design optimisation, enabling the STEP team to better understand and formulate the competing requirements of the FTOP.
We demonstrate that this can be achieved without increasing the number of JETTO runs required.
Taken together, these improvements significantly reduce the overall cost of iterating a design compared to previous approaches \textcite{marsden2022}.

Many plasma properties of interest vary from the core to the edge of the plasma as 1-D functions of radius (called \linebreak`profiles').
We demonstrate optimisation of the electron-cyclotron resonance heating (ECRH) deposition profile, with the goal of improving multiple properties of the safety factor profile, $q$.
To do so, we use \emph{multi-objective Bayesian optimisation} (MOBO), a powerful method for optimising costly black-box functions (described in \cref{sec:bo}) \cite{garnett2023}.
To our knowledge, this work is the first to use MOBO in shaping plasma profiles.

\section{Tokamak design optimisation}
We highlight a few key prior works in the field.
MOBO has previously been applied to the design of toroidal field coils for future fusion reactors, demonstrating improved performance compared to state-of-the-art multi-objective \linebreak genetic algorithms \cite{nunn2022}.
The optimisation of STEP current drive profiles using a scalar genetic algorithm was presented in \textcite{marsden2022}. 
Our paper extends this work by applying multi-objective Bayesian optimisation to the same problem, providing a comparison in terms of solution quality and informativeness. 
We also introduce more general parameterisations of the ECRH power density profile, and show that our method performs well despite the corresponding increase in difficulty of finding feasible solutions.

For a discussion of the plasma scenarios used in our simulations, see \textcite{meyer2022,kennedy2023}.
A comprehensive review of the STEP scenarios will be presented in a forthcoming paper.

\section{Multi-objective Bayesian optimisation of safety factor profiles for STEP}

In this section we present the $q$-profile design problem and formulate it as a multi-objective Bayesian optimisation task.

\subsection{Safety factor properties and ECRH parameterisation}

\begin{table}
\renewcommand{\arraystretch}{1.5}
\centering
\caption{Objective functions for the safety factor profile ($q$-profile)}
\label{qobjectives1}
\begin{tabular}{c>{\raggedright\arraybackslash}p{21ex}cc}
\toprule
ID & Desired property & Untransformed metric & Target \\
\midrule
1 & $q_0$ close to $q_\mathrm{min}$ & $\| q_0 - q_{\mathrm{min}} \|$ & $0$ \\
2 & $q_\mathrm{min}$ close to centre & $\hat{\rho}(q=q_{\mathrm{min}})$  & $0$ \\
3 & $q$ increasing &  $\frac{1}{N} \sum_{i=1}^N \mathbf{1}\Big(\Big[\frac{dq}{d\hat{\rho}}\Big]_i > 0\Big)$ & $1$ \\
4 & $2 < q_\mathrm{min} < 3$ & $q_\mathrm{min}$ & $[2, 3]$ \\
5 & $q=3$ towards edge & $\hat{\rho}(q=3)$ & $1$ \\
6 & $q=4$ towards edge & $\hat{\rho}(q=4)$ & $1$ \\
\bottomrule
\end{tabular}
\end{table}

\begin{figure*}[hb!]
    \centering
    \subfloat[Piecewise linear]{%
       \includegraphics[width=\columnwidth]{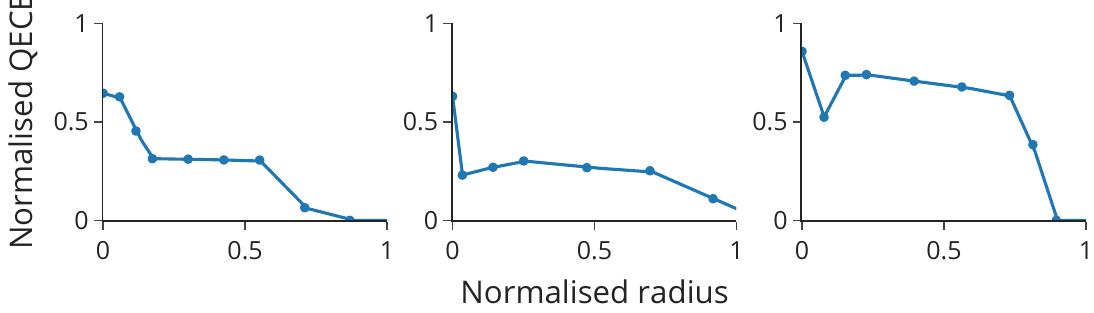}}
    \hfill
    \subfloat[Bézier]{%
       \includegraphics[width=\columnwidth]{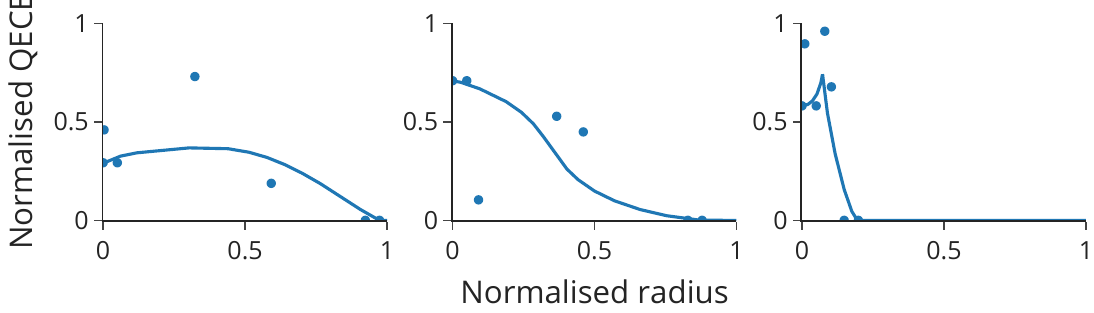}}
    \caption{Example random ECRH profiles (i.e., before any optimisation). `Piecewise linear' consists of 8 linear segments. `Bézier' is a single order 7 Bézier curve. Segment ends and Bézier control points are shown with markers.}
    \label{fig:example-parameterisations}
\end{figure*}

The shape of the $q$-profile plays an important role in improving the plasma confinement and minimising the impact of magnetohydrodynamic instabilities.
As STEP will operate non-inductively, ECRH is expected to be a primary actuator for shaping the safety factor profile via current drive \cite{meyer2022}.

The planned scenarios for STEP exhibit a monotonically increasing $q$-profile with $2 < q_{\mathrm{min}} < 3$ \cite{meyer2022}.
Monotonic $q$ profiles have been identified as contributing to increased stable normalised beta, $\beta_N$, \cite{menard2003} the avoidance of internal transport barriers \cite{connor2004} and reduced fast particle instabilities \cite{buttery2019}.
Maintaining a monotonic $q$ also means that the plamsa is operating in a regime where magnetohydrodynamic instabilities are generally better understood \cite{mahajan1982}.
Keeping $q > 2$ avoids the formation of the $(2,1)$ neoclassical tearing mode (NTM), allowing stable operation at high $\beta_N$ \cite{menard2004}.
Additional NTMs occur at rational values of $q$ \cite{walker2020}, and have been shown to exhibit improved stability if the rational $q$ surfaces are nearer the edge of the plasma \cite{maget2010} (although NTMs can also be destabilised by high magnetic shear, which is likely to occur at large minor radii).
Achieving these properties requires a large amount of auxiliary current drive towards the core of the plasma.

The exact mathematical representations of the objective functions have a significant impact on the solutions generated.
To quantify monotonicity, we compute the gradient of the $q$-profile at $N$ radial locations, and then find the fraction of locations for which the gradient is positive.
However, STEP FTOP candidates often have an undesirable region of reversed $q$ at the plasma centre, due to the vanishing of the bootstrap current on axis: we penalise this behaviour with two objectives that quantify the radial width and magnitude of reversed shear.
We set the target range for $q_\text{min}$ to be $2.2-2.5$.
We transform each of these desired $q$ properties using a custom function, which we call \textsc{SoftHat}, that rescales them to be maximisation objectives in $[0, 1]$. The function is detailed in \Cref{sec:softhat}.
The objectives are shown in \Cref{qobjectives1}, prior to transformation.

Note that our choice of objectives is not intended to be definitive: instead, the understanding obtained from optimisation against these heuristics will be used to narrow future high-resolution evaluations of the scenario space.

At present, we focus on purely theoretical design of ECRH power density profiles (which we refer to as `QECE').
We do not seek to design a launcher configuration or incorporate the physics of current drive.
Instead, we assume that any arbitrary profile can be achieved by the ECRH launchers, and define a general function parameterisation $g_\theta(\hat\rho)$ that can represent a wide variety of possible QECE profiles.

The first parameterisation we consider is a piecewise linear function with 12 parameters, introduced in \textcite{marsden2022}.
This function was designed to produce a limited range of plausible QECE profiles, with an on-axis peak and optional off-axis peak.

The second parameterisation is a Bézier curve, which is widely used in engineering design to represent complicated smooth shapes \cite{fitter2014}.
Bézier curves are bounded by the convex hull defined by the control points; hence, the curve can be constrained to be positive, which is difficult to achieve under other smooth parameterisations (e.g.\ cubic splines).

There is a tradeoff inherent in the number of parameters; using more parameters results in a representation that can produce a greater variety of profiles, but also makes it harder to find the optima.
We found that using 10 parameters for the Bézier curve was sufficient to replicate profiles generated by the piecewise linear parameterisation, while also being able to produce a diverse range of additional profiles that were not possible in the piecewise linear setting.

The resulting power density profile is scaled such that the total ECRH power is a predefined constant value.
Example profiles from each parameterisation are shown in \Cref{fig:example-parameterisations}.
See \Cref{sec:appendix-ecrh-parameterisations} for detailed mathematical definitions.

\subsection{Multi-objective optimisation}
The ECRH-$q$ optimisation problem is an example of a multi-objective optimisation (MOO) problem.
A common approach to MOO problems is to cast the multi-objective (vector) objective function $\boldsymbol{f}(\theta)$ to a single-objective (scalar) one by computing a weighted sum of the components, $f(\theta) = \boldsymbol{w}^T \boldsymbol{f}(\theta)$.
However, using a weighted sum introduces preconceptions about the desired balance between the objectives, and the optimisation routine will only produce solutions that align with the pre-determined tradeoff.
Moreover, tuning the weights $\boldsymbol{w}$ is resource-intensive, and relies on human intuition or expertise.

Instead, taking a `true' multi-objective (vector) approach allows us to gain understanding of how the objectives interact and conflict, and facilitates reasoned decision-making after observing a wide variety of possible solutions to the tradeoff.
To perform true multi-objective analysis, MOO algorithms seek to find a representative set of \textit{Pareto optimal} solutions.
More precisely, an ECRH parameter vector $\theta$ is \emph{Pareto-optimal} if there is no other vector that is (i) at least as good as $\theta$ in all objectives and (ii) better than $\theta$ in at least one objective.
Vectors that are worse than $\theta$ in all objectives are \emph{Pareto-dominated} by $\theta$, and the set of all Pareto-optimal solutions is known as the \textit{Pareto front}.
Solutions on the Pareto front therefore reflect the consequences of assigning different priorities to the objectives.

\subsection{Bayesian optimisation}
\label{sec:bo}
Bayesian optimisation (BO) is a principled approach to performing gradient-free global optimisation of costly black-box functions.
We provide a brief overview of the method as applied to our problem; for a detailed tutorial on BO, see \textcite{garnett2023}.

Global optimisation tasks involve a compromise between \textit{exploitation} and \textit{exploration}: the optimiser must decide whether to evaluate trial solutions that are close to its current best guess of the optimal solution (`exploitation') or try solutions from an as-yet unobserved region (`exploration').
Many global optimisation algorithms tackle this dilemma using stochastic exploration; for example, in a genetic algorithm, random mutations are introduced in every generation.
Bayesian optimisation instead works by building a probabilistic model of the mapping from inputs $\theta$ to objectives $\boldsymbol{f}(\theta)$, and using the model's uncertainty predictions to resolve the explore-exploit tradeoff.

Firstly, a probabilistic model is fit to all of the previously observed input-objective pairs, $\mathcal{D} = \left\{(\theta_i, \boldsymbol{f}(\theta_i)\right\}_{i=1}^n$, by maximising the marginal likelihood.\footnote{An initial set of observations is generated by selecting inputs using random or pseudo-random sampling and generating the corresponding objective values.}
The model used is a \textit{Gaussian process} (GP).
A GP is an extension of the Gaussian distribution to infinite dimensions, and can be understood as representing a probability distribution over functions \cite{williams2006}.

\begin{definition}{Gaussian process.}
A Gaussian process over functions $f: \mathcal{X} \to \mathbb{R}$ is defined by a mean function ${m: \mathcal{X} \to \mathbb{R}}$ and a positive semi-definite covariance function ${k: \mathcal{X} \times \mathcal{X} \to \mathbb{R}}$ such that $k(\cdot, \cdot) \geq 0$. A GP is denoted by
\begin{equation}
f \sim \mathcal{GP}\left(m, k\right).
\end{equation}
Then, the values of $f$ at a finite set of evaluation points $\mathbf{x} = \left[x_1, \dots, x_N\right]\subset \mathcal{X}$ are distributed according to
\begin{equation}
f(\mathbf{x}) \sim \mathcal{N}(\boldsymbol{\mu}, \boldsymbol{\Sigma}), 
\end{equation}
where the vector $\boldsymbol{\mu} = \left[m(x_1), m(x_2), \dots\right]^T$ and the matrix $\boldsymbol{\Sigma}$ has elements $\left[\boldsymbol{\Sigma}\right]_{ij} = k(x_i, x_j)\ \forall i, j \in \{1,\dots,N\}$.
\end{definition}

In our example problem, for any candidate ECRH parameter vector $\theta$, the GP provides a predictive probability distribution for the vector of objective values $\boldsymbol{f}(\theta)$ given in \Cref{qobjectives1}.

To select the next points to trial, BO uses the predictive model in conjunction with an \emph{acquisition function}, $\alpha : \Theta \to \mathbb{R}$, which estimates the `usefulness' of observing the objective value of a previously unseen candidate $\theta \in \Theta$ based on the GP predictions.
The acquisition function is cheaper to evaluate than the true objective function, so gradient ascent can be performed on $\alpha$ to find the candidate solution(s) that are most useful to evaluate next.
The values of the objective functions for the candidate(s) are then evaluated (in our case, by running JETTO to generate a $q$-profile, and computing the objective values from the $q$-profile properties), and the model is updated to account for the new observation.

The choice of acquisition function determines the approach to resolving the explore-exploit tradeoff.
We chose batch Noisy Expected Hypervolume Improvement (qNEHVI), which is currently a state-of-the-art acquisition function for multi-objective optimisation \cite{daulton2021}.
The hypervolume of an estimated Pareto set with $N$ objectives, $\mathrm{HV}[\mathcal{P}]$, is the $N$-dimensional Lebesgue integral of the objective space dominated by $\mathcal{P}$.
At optimisation iteration $n$, given an estimate of the Pareto front, $\mathcal{P}_n$, and a posterior GP distribution over objective values, $p(\boldsymbol{f}(\theta) | \mathcal{D})$, the expected noisy hypervolume improvement corresponding to observing a candidate $\theta$ is given by
\begin{align}
    \alpha_\mathrm{NEHVI} = \mathbb{E}_{\boldsymbol{f}} \left[\mathrm{HV}(\mathcal{P}_n \cup \theta) - \mathrm{HV}(\mathcal{P}_n)\right]
\end{align}
where the expectation is taken with respect to the GP posterior distribution.
In qNEHVI, the single input $\theta$ is replaced with a set of $m$ inputs $[\theta_1, \dots, \theta_m]$.
In practice, the expectation and the integrals are approximated by Monte Carlo sampling.

\subsection{Experimental setup}
For our experiments, we fixed a budget of 8 evaluation steps, each with 32 parallel JETTO evaluations. 
This corresponds to approximately 24 hours of wall-clock time, and is the same budget used for the scalar genetic algorithm method in \textcite{marsden2022}.

We use BoTorch \cite{botorch} for Bayesian optimisation.
We use an independent GP model with a Matérn-$\frac{5}{2}$ kernel for each output, using the default BoTorch hyperparameter priors.
We normalise the ECRH parameter vectors to $[0, 1]$ and standardise the outputs to have zero mean and unit variance.
These are standard choices in Bayesian optimisation, which we selected for their simplicity.
Note that a Matérn-$\frac{5}{2}$ kernel models less-smooth functions than, say, a squared exponential kernel, leading to a model that is slightly more robust.
We generate batches of $32$ candidate solutions for each step.
We found that with this batch size it was best to perform sequentially greedy optimisation of the acquisition function; this ensured the cost of candidate generation was small, so that the majority of the compute budget could be spent on JETTO evaluations.
We generated 32 initial candidates using scrambled Sobol sampling.

Our code is available on GitHub\footnote{\textsc{\url{https://github.com/theo-brown/jetto-mobo}}} as a flexible Python package for using Bayesian optimisation with JETTO.

\begin{figure*}
    \centering
    \begin{tikzpicture}
        \node[anchor=center,rotate=90] at (0,0) {\parbox{20ex}{\subcaption{Piecewise linear}\label{fig:pl_mobo_solutions}}};
        \node[anchor=west] at (1,0) {\includegraphics[width=0.8\textwidth, trim=0 0 0 70, clip]{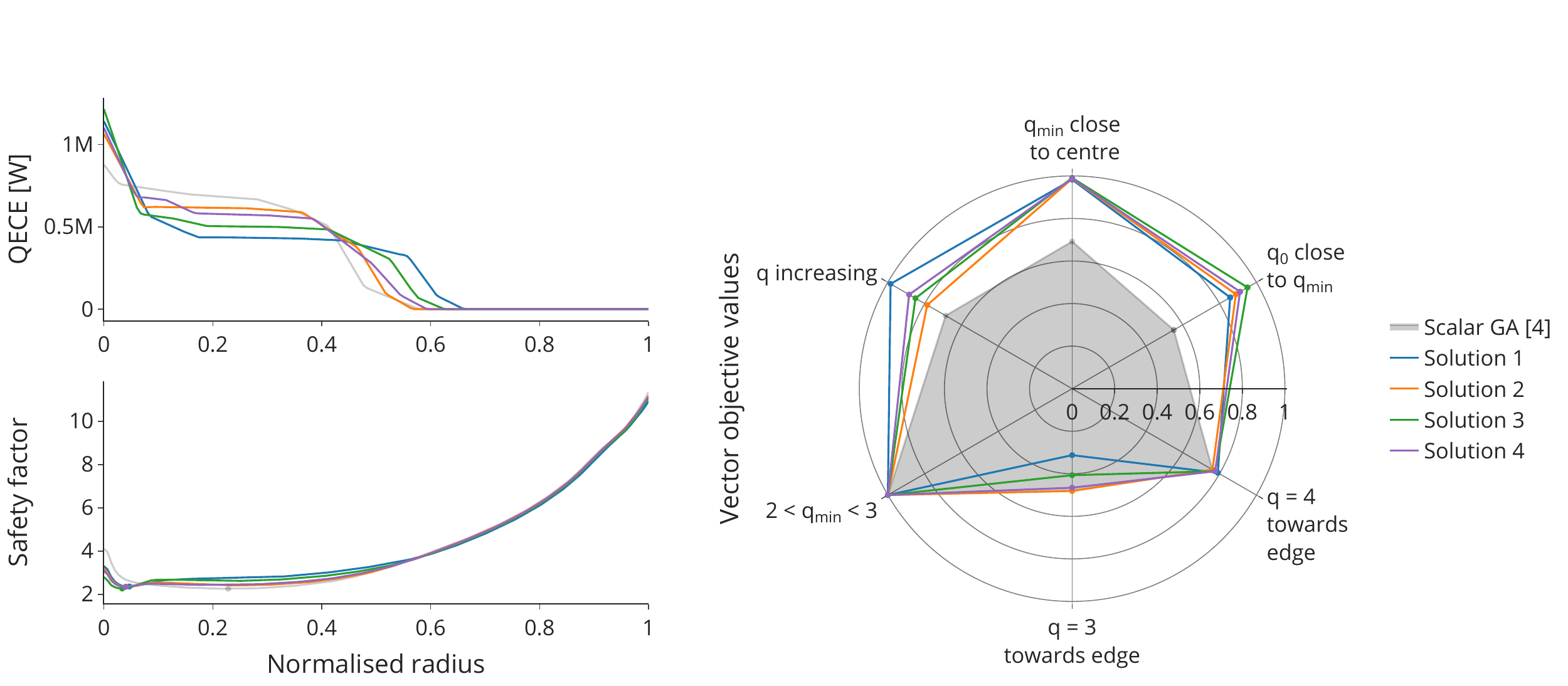}};
    \end{tikzpicture}
    \begin{tikzpicture}
        \node[anchor=center,rotate=90] at (0,0) {\parbox{20ex}{\subcaption{Bézier}\label{fig:bz_mobo_solutions}}};
        \node[anchor=west] at (1,0) {\includegraphics[width=0.8\textwidth, trim=0 0 0 70, clip]{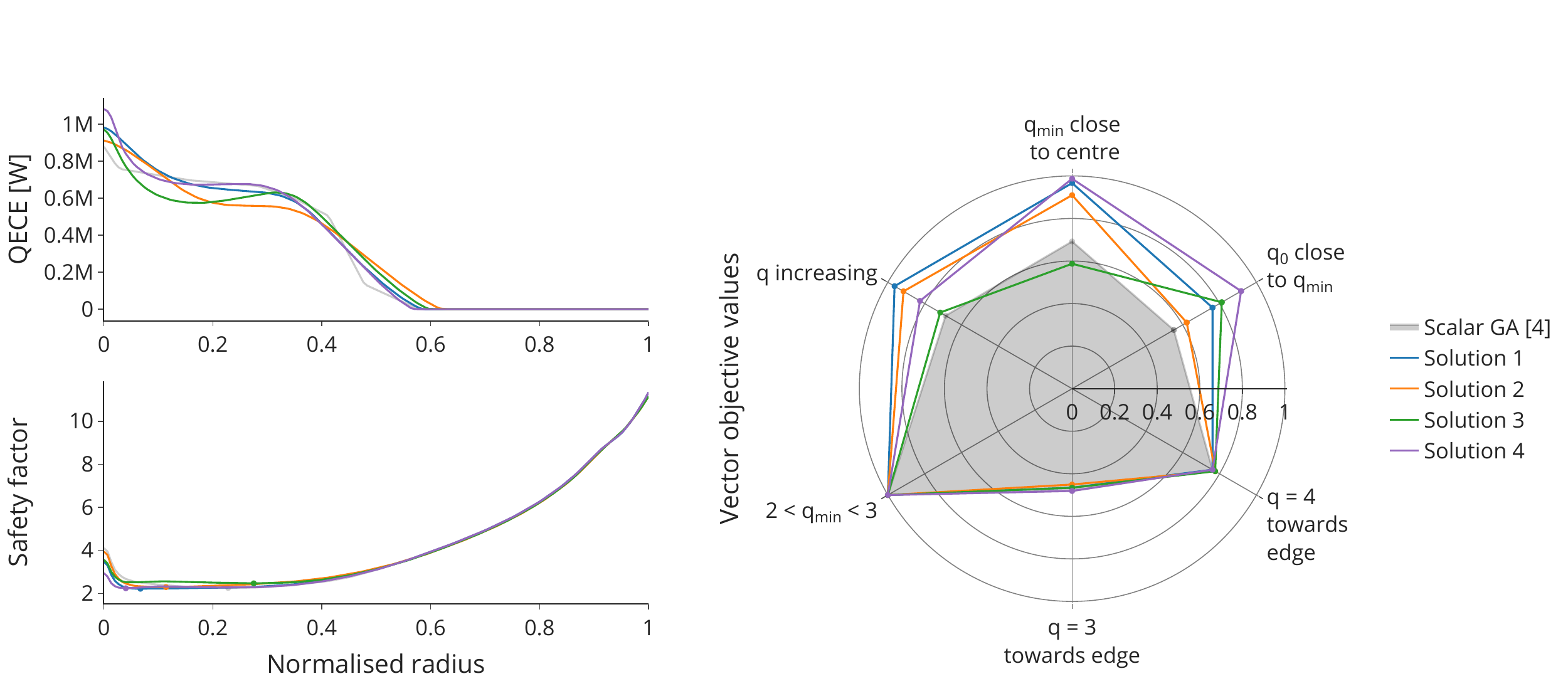}};
    \end{tikzpicture}
    
    \caption{A subset of the solutions to the ECRH-$q$ optimisation problem generated by multi-objective BO, compared to the solution from a scalar genetic algorithm (GA) with piecewise linear ECRH (from \cite{marsden2022}). In each case, the GA's solution is Pareto dominated by our solutions. In addition, multi-objective BO identifies many other optimal solutions, highlighting potential gains from different weighting of the objectives.}
    \label{fig:mobo_solutions}
\end{figure*}

\section{Results}

Our results are shown in \Cref{fig:mobo_solutions}, alongside the original genetic algorithm (GA) solution from \textcite{marsden2022}. 
We have manually selected a representative subset of the Pareto-optimal solutions for each parameterisation for illustrative purposes; our MOBO algorithm generated 8 and 10 solutions for the piecewise linear and Bézier parameterisations respectively.

\begin{figure}
    \centering
    \includegraphics[width=0.9\columnwidth]{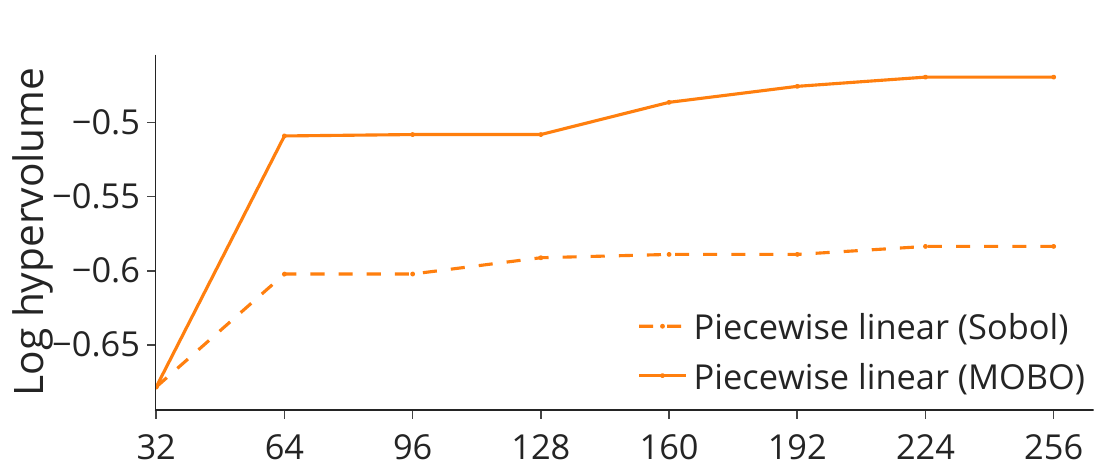}
    \includegraphics[width=0.9\columnwidth]{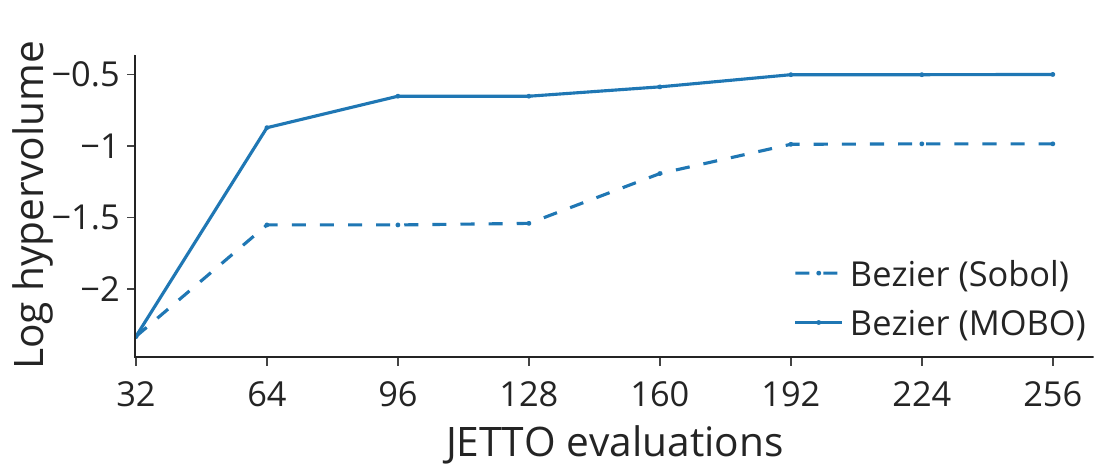}
    \caption{Progress graphs showing the log hypervolume at each iteration step. In both cases, multi-objective BO converges significantly faster than Sobol sampling.}
    \label{fig:convergence}
\end{figure}

\autoref{fig:convergence} displays the progress of the optimiser. As the log hypervolume reached a plateau, we conclude that the compute budge was sufficient for the algorithm to converge.

We highlight a few key conclusions:
\begin{enumerate}
    \item[(1)] Under both parameterisations, the GA solution is Pareto dominated by our solutions: that is, our method generates solutions that are \emph{better in every objective} than the GA.
    \item[(2)] The Bézier representation is able to produce competitive solutions, despite incorporating no prior knowledge; in contrast, the piecewise linear parameterisation was hand-crafted and constrained to produce plausible profiles. This highlights the power of the MOBO approach, as it can optimise a general parameterisation from scratch.
    \item[(3)] Our results highlight the tradeoffs inherent in this task. For example, solution 3 in \cref{fig:pl_mobo_solutions} and solution 4 in \cref{fig:bz_mobo_solutions} perform the best at the core ($q_0$ close to $q_\mathrm{min}$ and $q_\mathrm{min}$ close to centre) at the expense of reduced global monotonicity ($q$ increasing). The existence of such compromises is expected, as we are optimising the allocation of a fixed amount of EC power.
\end{enumerate}

\section{Discussion and further work}

We now discuss some of the challenges encountered, areas for improvement, and overall takeaways.

\emph{Dealing with failures.}
Many JETTO runs fail to converge, which presents a problem for optimisation routines.
If the optimiser simply discards failures, the optimiser may repeatedly resample at or near the failure point, resulting in wasted many wasted evaluations.
Replacing the failed runs with an artificial objective value (`imputation') can avoid this problem, but can reduce the model's ability to generalise and introduces an additional hyperparameter to tune.
In our case, we performed a grid search to find the best value, which we found to be 0.3.
A more principled approach would involve developing new acquisition functions that can handle failures, which is a current area of active research \cite{iwazaki2023failure}.

\emph{ECRH parameterisations.}
One of the challenges in designing an ECRH parameterisation is to ensure sufficient flexibility without making the search space too large. 
For example, we found that an unconstrained piecewise linear function was overly expressive, which meant that it was difficult to find `good' regions of parameter space.
It would be illuminating to perform studies with ECRH parameterisations that better reflect realistic heating configurations, such as Gaussian bumps representing the deposition from EC beam launchers; this would allow optimisation of machine-relevant scenarios.

\emph{Many-objective optimisation.}
In general, multi-objective optimisation algorithms work best with a small number of objectives, and can exhibit significant decay in performance in so-called \emph{many}-objective problems.
Strategies for tackling many-objective problems include performing dimensionality reduction to find a smaller set of uncorrelated objectives, guiding the search algorithm by introducing objectives sequentially, and scalarisation methods \cite{fleming2005, kasimbeyli2019}.
Some of the objectives that we selected could be reformulated as constraints, which could also improve performance by reducing the number of objectives; for example, the $2 < q_{\mathrm{min}} < 3$ objective could be treated as a constraint, as any profiles with $q < 2$ will not be considered for STEP.
Finally, alternative acquisition functions such as variants of Thompson sampling \cite{kandasamy2018} might scale better to high-dimensional tokamak design problems.

Extensions such as these are a promising avenue for future work, and could enable multi-objective optimisation of a greater number of plasma properties.

\section{Conclusion}
Using multi-objective Bayesian optimisation, we find a varied range of ECRH heating profiles that represent optimal tradeoffs between the desired properties of the $q$-profile for STEP, without any increase in compute requirements.
With the same number of evaluations, our method generates solutions that outperform those generated with a genetic algorithm operating on a scalarised objective.
Moreover, our solutions are widely spaced along the Pareto front and hence represent diverse tradeoffs; in contrast, the genetic algorithm's solutions are clustered around a specific tradeoff, determined by the scalar weighting of the objectives (see \textcite{marsden2022}).
The necessity of balancing competing objectives is inherent in multi-objective problems; we hope our work encourages fusion scientists and engineers to adopt multi-objective Bayesian optimisation to facilitate informed decision making when designing future devices.

\printbibliography

\clearpage
\appendices

\begin{figure*}
    \centering
    \subfloat[Maximisation objective ($x_{\textsc{lower}}=0.5,\allowbreak \textsc{start}=0.8, \textsc{end}=1$)]{%
       \includegraphics[width=0.3\textwidth]{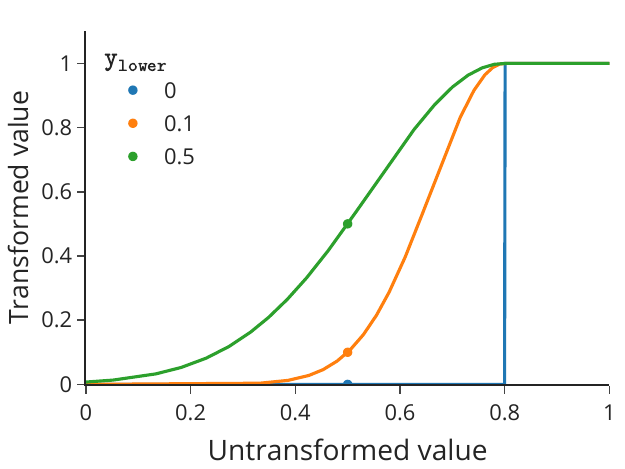}}
    \hfill
    \subfloat[Minimisation objective ($\textsc{start} = 0,\allowbreak {\textsc{end} = 0.2}, x_{\textsc{upper}}=0.5$)]{%
       \includegraphics[width=0.3\textwidth]{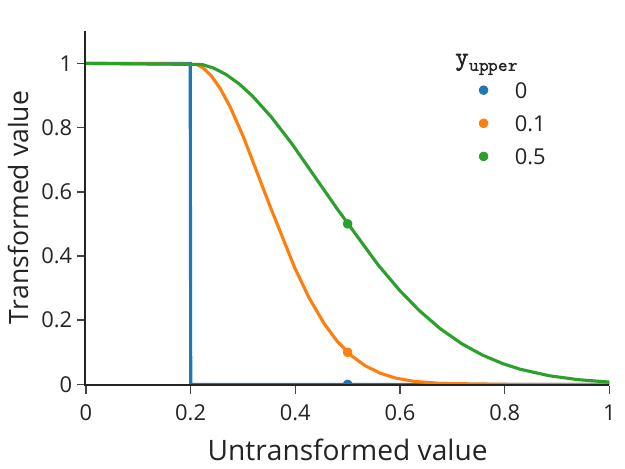}}
    \hfill
    \subfloat[Target range objective, $x_{\textsc{lower}}=0.2,\allowbreak \textsc{start}=0.6,\allowbreak \textsc{end}=0.7,\allowbreak x_{\textsc{upper}}=0.8$]{%
       \includegraphics[width=0.3\textwidth]{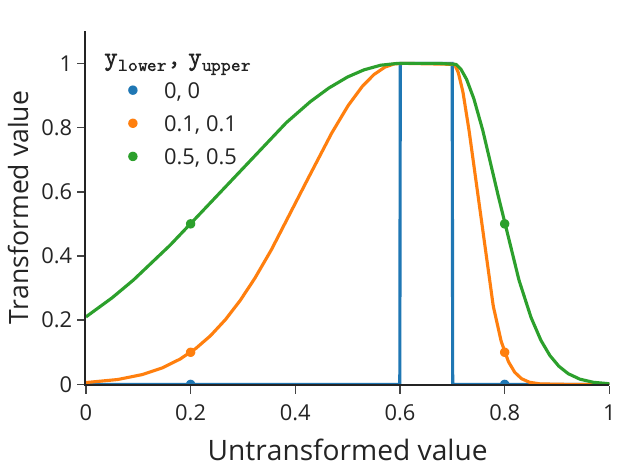}}
    \caption{Demonstration of the $\textsc{SoftHat}$ transformation function. The function first increases as a squared exponential, passing through $(x_{\textsc{lower}}, y_{\textsc{lower}})$, then is uniformly $1$ between \textsc{start} and \textsc{end}, before decreasing as a squared exponential, passing through $(x_{\textsc{upper}}, y_{\textsc{upper}})$. Setting $y_{\textsc{lower}}$ or $y_{\textsc{upper}}$ to $0$ recovers a square pulse (`top-hat') function.}
\end{figure*}

\section{SoftHat transformation}
\label[apdx]{sec:softhat}
We defined a transformation to scale the objective values to $[0,1]$, where $1$ is the `best' value.
Our key requirements were:
\begin{itemize}
    \item Easy to interpret
    \item Smooth
    \item Can represent maximisation and minimisation objectives
    \item Can represent objectives where the desired value is within a given range
\end{itemize}

Our transformation is:
\begin{align*}
&\textsc{SoftHat}\left(x, x_{\textsc{lower}}, y_{\textsc{lower}}, \textsc{start}, \textsc{end}, x_{\textsc{upper}}, y_{\textsc{upper}}\right)
\nonumber \\ & = 
\begin{cases}
\exp\left(-k_{\textsc{lower}}\left(x - \textsc{start}\right)^2\right) & x < \textsc{start} \\
1 & \textsc{start} \leq x \leq \textsc{end} \\
\exp\left(-k_{\textsc{upper}}\left(x - \textsc{end}\right)^2\right) & x > \textsc{end}\mathrm{,} \\
\end{cases}
\end{align*}
where
\begin{equation*}
    k_{\textsc{lower}} = -\frac{\log(y_{\textsc{lower}})}{\left(x_{\textsc{lower}} - \textsc{start}\right)^2}
\end{equation*}
\begin{equation*}
    k_{\textsc{upper}} = -\frac{\log(y_{\textsc{upper}})}{\left(x_{\textsc{upper}} - \textsc{end}\right)^2}\mathrm{.}
\end{equation*}
This choice of the coefficients sets the rate of decay such that the function passes through  $(x_{\textsc{lower}}, y_{\textsc{lower}})$ and $(x_{\textsc{upper}}, y_{\textsc{upper}})$.

\begin{table*}[b!]
\begin{minipage}{1.2\columnwidth}
\caption{Parameter bounds for the piecewise linear parameterisation.}
\label{tab:piecewiselinearbounds}
\begin{tabular}{clcc}
\toprule
Parameter & Label & Lower & Upper \\
\midrule
$p_0$ & Normalised on-axis power & $0$ & $1$ \\
$p_1$ & Normalised radius of $b$ & $0$ & $0.1$ \\
$p_2$ & Normalised power at $b$ as a fraction of on-axis power & $0$ & $1$ \\
$p_3$ & Normalised power fraction halfway between $b$ and $d$ & $0$ & $1$ \\
$p_4$ & Normalised radius of $d$  & $0.1$ & $0.3$\\
$p_5$ & Normalised power at $d$ & $0$ & $1$ \\
$p_6$ & Normalised power fraction $\frac{1}{3}$ between $d$ and $g$ & $0$ & $1$ \\
$p_7$ & Normalised power fraction $\frac{2}{3}$ between $d$ and $g$ & $0$ & $1$\\
$p_8$ & Normalised radial distance between $d$ and $g$ & $0.1$ & $0.8$ \\
$p_9$ & Normalised power at $g$ & $0$ & $1$\\
$p_{10}$ & Normalised power halfway between $g$ and turnoff ($i$) & $0$ & $1$ \\
$p_{11}$ & Normalised radial distance between $g$ and turnoff & $0$ & $0.5$ \\
\bottomrule
\end{tabular}
\end{minipage}%
\hfill
\begin{minipage}{0.75\columnwidth}
\centering
\caption{Relationship between parameters and node \\locations for the piecewise  \\ linear parameterisation.}
\label{tab:piecewiselinearparameters}
\begin{tabular}{cccc}
\toprule
Node & $x$ & $y$ \\
\midrule
$a$ & $0$ & $p_0$ \\
$b$ & $p_1$ & $p_2 a_y$ \\
$c$ & $\frac{b_x + d_x}{2}$ & $p_3 b_y + (1-p_3) d_y$ \\
$d$ & $p_4$ & $p_5$ \\
$e$ & $\frac{2}{3} d_x + \frac{1}{3} g_x$ & $p_7 f_y + (1-p_7)d_y$ \\
$f$ & $\frac{1}{3} d_x + \frac{2}{3} g_x$ & $p_6 g_y + (1-p_6) d_y$ \\
$g$ & $d_x + p_8$ & $p_9$ \\
$h$ & $\frac{g_x + i_x}{2}$ & $p_{10} g_y$ \\
$i$ & $g_x + p_{11}$ & $0$ \\
\bottomrule
\end{tabular}
\end{minipage}
\end{table*}

\newpage
\section{ECRH power density profile parameterisations}
\label[apdx]{sec:appendix-ecrh-parameterisations}
\subsection{Piecewise linear}
The piecewise linear function parameterisation was designed to permit a variety of shapes; in particular, it can create a) a sharp peak on axis to prevent the formation of a current hole, and either b) a trough next to the on-axis peak and create an off-axis peak, or c) monotonically decreasing profiles with no trough.
The function is parameterised by 12 parameters (\Cref{tab:piecewiselinearbounds}), defining 9 nodes (\Cref{tab:piecewiselinearparameters}).
In \Cref{tab:piecewiselinearparameters}, components $x$ and $y$ are the node's position in normalised radius and normalised ECRH power density respectively.
A straight line is interpolated between each of the adjacent pairs of nodes to form the normalised ECRH power density profile.

This formulation ensures that the nodes remain in the intended order.
In addition, defining the nodes in relation to one another means that adjusting the radial position of the off-axis peak does not affect its overall shape.
\subsection{Bézier}
The 10-dimensional parameter vector $\left(p_0, \dots, p_9\right)$ is combined with two smoothing hyperparameters $\delta_0$ and $\delta_1$, which adjust the gradient of the curve at the end points, to produce the control point vector $\left((0, p_0), (\delta_0, p_0), (p_1, p_2), \dots, (p_7, p_8), (p_9 - \delta_1, 0), (p_9, 0)\right)$.
Note that this constrains the profile to finish at 0.

\end{document}